\documentclass[useAMS,usenatbib]{mn2e}
  \def\pks{PKS~0558--504}

  \def\xmm{{\it XMM-Newton}}

  \def\swift{{\it SWIFT}} 
  \def\rxte{{\it RXTE}}

  \def\ltsima{$\; \buildrel < \over \sim \;$}
  \def\simlt{\lower.5ex\hbox{\ltsima}} 
  \def\gtsima{$\; \buildrel > \over \sim \;$}
  \def\simgt{\lower.5ex\hbox{\gtsima}} 
  \def\plm{$\pm$}
\usepackage{graphics,graphicx}
\usepackage{epsfig}
\usepackage{dcolumn}
\usepackage{rotating}

\title[Long-term monitoring of PKS~0558--504 with Swift]{Long-term monitoring of PKS~0558--504 with Swift: the disc-corona link}
\author[M. Gliozzi et al.]{M. Gliozzi,$^{1}$\thanks{E-mail: mgliozzi@gmu.edu},
I.E. Papadakis,$^{2,3}$ D. Grupe,$^{4}$ W.P. Brinkmann,$^{5}$ and C. R\"ath$^{5}$\\
$^{1}$ School of Physics, Astronomy and Computational Sciences,
George Mason University, 4400 University Drive, Fairfax, VA 22030\\
$^{2}$ Department of Physics and Institute of Theoretical and Computational Physics, University of Crete, 71003 Heraklion, Greece\\
$^{3}$ IESL, Foundation for Research and Technology, 71110 Heraklion, Greece\\
$^{4}$ Department of Astronomy and Astrophysics, Pennsylvania State University, 525 Davey Lab, University Park, PA 16802, USA\\
$^{5}$ Max-Planck-Institut f\"ur 
extraterrestrische Physik, Postfach 1312, D-85741 Garching, Germany}
\begin{document}
%
%
%
\maketitle
\begin{abstract}
PKS 0558­-504 is a highly variable, X-ray bright, radio-loud, Narrow-Line Seyfert 
1 galaxy with super-Eddington accretion rate and extended jets that do
not dominate the emission beyond the radio band. Therefore, this source represents 
an ideal laboratory to shed some light on the central engine in highly accreting
systems and specifically on the link between accretion disc and  corona. Here we present the results from a 1.5 year monitoring with \swift\ XRT 
and UVOT. The simultaneous coverage at several wavelengths confirms that PKS 0558­-504 
is highly variable in any band from optical, to UV and X-rays, with the latter
showing the largest amplitude changes but with the UV emission dominating the radiative 
output. A cross-correlation analysis reveals a tight link between the emission
in the optical and UV bands and provides suggestive evidence in favor of a scenario
where the variability originates in the outer part of the accretion flow and 
propagates inwards before triggering the activity of the X-ray emitting corona.
Finally, a positive correlation between the soft X-ray flux 
and the hard photon index suggests that in PKS 0558­-504 the seed photons are
provided to the corona by the soft excess component.
\end{abstract}
\section{Introduction}
More than three decades of multi-wavelength observations of active galactic 
nuclei (AGN) have revealed that these objects are powerful emitters  of variable
radiation from the radio band to $\gamma$-ray energies and that 
most of the radiation is emitted in the optical-UV band and in the X-rays.
In the current leading scenario, AGN are powered by accretion onto a central 
supermassive black hole, the optical-UV emission is thermal radiation 
produced (directly and/or via reprocessing) from the accretion flow, and the X-rays are 
produced through Comptonisation process in a putative corona. While this general picture is  
widely accepted, the details concerning the interaction between disc and corona as well
as the geometry and physical state of the latter (for example, thermal static vs. ``cool"
dynamic medium) and the origin of the variability in different bands (e.g., intrinsic
disc emission variation vs. reprocessing of the X-ray emission)
are still poorly understood.

One of the best way to address these outstanding open questions is to investigate the
long-term variability in several different energy bands that track the activity of the main components 
of the central engine, namely the disc and the corona. In recent years, several multi-wavelength
 variability studies have confirmed the validity of this approach and revealed in some
cases the existence of coordinated variability patterns and in others the lack
of coordinated changes \citep[e.g.,][]{maoz00,shem01,utt03,arev08,bree09,arev09,bree10}. This apparent
discrepancy of long-term behaviour in different AGN can be partly ascribed to the
lack of simultaneous coverage and to the 
difficulty of adequately capturing the disc activity using 
ground-based optical telescopes that are often hampered by weather conditions. However, it  
could also be due to intrinsic differences in some fundamental properties of the central engine, 
such as the BH mass, $M_{\rm BH}$, and the accretion rate in Eddington units, $\dot m$ \citep[e.g.,][]{utt03}.

In order to shed further light on the link between disc and corona, 
it is crucial to simultaneously study in different energy bands 
the activity of variable black hole systems whose  $M_{\rm BH}$ 
and $\dot m$ are well constrained. The advent of \swift\ with its 
flexibility coupled with the simultaneous coverage of several bands in the 
optical/UV and in the X-rays offers the ideal tool for this kind of studies.

Over the last few years, we have carried out a multi-wavelength  campaign of the 
X-ray bright radio-loud Narrow-line Seyfert 1 (NLS1) \pks\ with \rxte, the ATCA, 
\xmm, and with \swift. The long-term \rxte\ monitoring indicates that
 \pks\ is highly variable on timescales ranging from minutes to years with
frequent flares and continuous  ``achromatic" large-amplitude variations
that are similar to the behaviour shown by Galactic black holes in 
highly-accreting intermediate spectral states \citep{glioz07}. 
The radio observations reveal the presence of aligned extended structures
suggesting the presence of bipolar jets with a relatively large viewing angle \citep{glioz10}.
A deep \xmm\ observation (5 consecutive orbits) indicates that \pks\
lacks any significant intrinsic absorption and strong reflection features;
the source has X-ray spectral and temporal
variability typical of radio-quiet Seyfert galaxies, and has a highly variable
soft excess well described by a low-temperature Comptonisation model \citep{papa10a,papa10b}.
A 10-day monitoring campaign in September 2008 with \swift\ combined with simultaneous
\xmm\ and archival radio data constrains the black hole mass to 
$M_{\rm BH}=(2-4)\times10^8~{M_\odot}$
confirming that \pks\ is accreting at or above the Eddington limit. The \swift\
observations also 
reveal that the spectral energy distribution (SED) is dominated by 
the optical-UV emission, and that the jet appears to be  relevant  
only in the radio band \citep{glioz10}. 

Based on these findings, we can conclude that 
\pks\ is a ``clean" system accreting at super-Eddington rate, ideal to study 
the central engine and specifically the disc-corona interaction
in the high-accretion regime. For this purpose,
we investigate the multi-wavelength
behaviour of \pks\ as seen by \swift\ XRT and UVOT over a time interval
of one and a half years. The outline of the paper is as follows. First,  
in Section 2, we describe the observations and data reduction. In Section 3 
we study the long-term flux variability in the X-ray band and in six
UVOT filters, whereas in Section 4 we perform an inter-band correlation analysis. 
In Section 5, we carry out a spectral analysis
of the broadband SED and investigate the link between disc, corona and soft excess.
Finally, in Section 6 we summarize the main results and discuss their implications.
Hereafter, we adopt a cosmology with $H_0=71{\rm~km~s^{-1}~Mpc^{-1}}$,
$\Omega_\Lambda=0.73$ and $\Omega_{\rm M}=0.27$ \citep{ben03}. 
With the assumed cosmological parameters, the luminosity
distance of \pks\  is 642 Mpc ($z=0.137$).

\section{Observations and Data Reduction}
\pks\ was observed by \swift\ \citep{ger04} between 9 September 
2008 and 30 March 2010 with one pointing per week (with typical exposure
per visit of $\sim$2ks), with the exception of
the first ten days when the source was observed on daily basis
\citep{glioz10}. The details of the \swift\ monitoring campaign
(dates, exposures, X-ray count rates, and UVOT
magnitudes) are summarized in Tables 1 and 2, where only the first 
three entries are shown for illustrative purposes. The complete tables are  
available in electronic format.

The \swift\ XRT \citep{burro05} 
observations were performed in windowed timing mode  
to avoid possible pile-up effects (\citealt{hill04}). The 
event file of the observation was created by using the Swift analysis tool 
xrtpipeline version 0.11.4. Source photons were extracted from a box with length 
radius of 40 pixels centered on the source; the background was selected 
from a similar source-free region. Only single to 
quadruple events in the energy range of 0.3--10 keV were selected for further
 analysis. Source and background spectra were extracted from the event file 
by using XSELECT version 2.3. Spectra were rebinned within grppha 3.0.0 to 
have at least 20 photons per bin. The auxiliary response files were created 
by the Swift tool {\tt xrtmkarf}. We used the response matrix version 012 
with a grade selection 0-12. 

In addition to the X-ray data, we also obtained photometry with the UV/Optical 
Telescope (UVOT; \citealt{rom05}) in the V, B, U, UVW1, UVM2, and UVW2 
filters. Source photons were 
extracted from a circular region with r = 5\arcsec, and the background from an 
annulus around the source with an inner radius of 7\arcsec\ and an outer 
radius of 20\arcsec. The UVOT tool {\tt uvotsource} was used to determine the 
magnitudes and fluxes \citep{pool08,breev10}.  The fluxes were corrected for Galactic reddening
($E_{\rm B-V}$=0.044; \citealt{schle98}) with the standard
reddening correction curves by  \citet{card89} as described by equation 2 in
\citet{rom09}.

\setcounter{table}{0}
\begin{table*} 
\centering
\begin{minipage}{180mm}
\caption{Observation log of PKS 0558-504}
\begin{tabular}{lcccccccccc} 
\hline        
\hline
\noalign{\smallskip} 
Segment & Start time (UT) & End Time (UT) & MJD & \multicolumn{7}{c}{Observing time given in s}\\
 & & & & $\rm T_{\rm XRT}$& $\rm T_{\rm V}$ & $\rm T_{\rm B}$ &$\rm T_{\rm U}$ & $\rm T_{\rm UVW1}$ &
$\rm T_{\rm UVM2}$ & $\rm T_{\rm UVW2}$\\
\noalign{\smallskip}
\hline
\noalign{\smallskip}
011 & 2008-09-18 17:18 & 2008-09-18 20:42 & 54727.79 & 2056 & 163 & 163 & 163 & 326 & 490 & 652 \\
012 & 2008-09-25 00:08 & 2008-09-25 18:14 & 54734.38 & 1884 & 147 & 147 & 147 & 294 & 404 & 590 \\
013 & 2008-10-02 00:52 & 2008-10-02 13:51 & 54741.31 & 1516 & 127 & 126 & 127 & 253 & 296 & 507 \\
\noalign{\smallskip}
\hline        
\hline
\end{tabular}
\end{minipage}
\label{tab1}
\end{table*} 
 
\setcounter{table}{1}
\begin{table*} 
\begin{minipage}{180mm}
\caption{\swift~XRT count rates and hardness ratios and UVOT Magnitudes of PKS 0558-504}
\begin{tabular}{lcccccccc}
\hline        
\hline
\noalign{\smallskip}
Segment & XRT rate & XRT HR & V & B & U & UVW1 & UVM2 & UVW2 \\
\noalign{\smallskip}
\hline
\noalign{\smallskip}
011 & 1.37\plm0.03 & --0.03\plm0.02  & 14.83\plm0.02 & 15.03\plm0.01 & 13.84\plm0.01 & 13.55\plm0.01 & 13.25\plm0.01 & 13.29\plm0.01 \\
012 & 1.20\plm0.03 & --0.05\plm0.02  & 14.90\plm0.03 & 15.07\plm0.01 & 13.85\plm0.01 & 13.58\plm0.01 & 13.31\plm0.01 & 13.38\plm0.01 \\
013 & 1.87\plm0.03 & --0.08\plm0.02  & 14.86\plm0.03 & 15.04\plm0.02 & 13.86\plm0.01 & 13.55\plm0.01 & 13.22\plm0.01 & 13.33\plm0.01 \\
\noalign{\smallskip}
\hline        
\hline
\end{tabular}
\end{minipage}
\label{tab2}
\vskip 0.1cm
\begin{flushleft}
The magnitudes were corrected for reddening with $E_{\rm B-V}$=0.044.
The errors given in this table are statistical errors.\\
Note: The complete tables are available in the on-line version.
\end{flushleft}
\end{table*}
\section{Variability analysis of \pks}
\begin{figure}\
\begin{center}
\includegraphics[bb=80 25 470 570,clip=,angle=0,width=8.5cm]{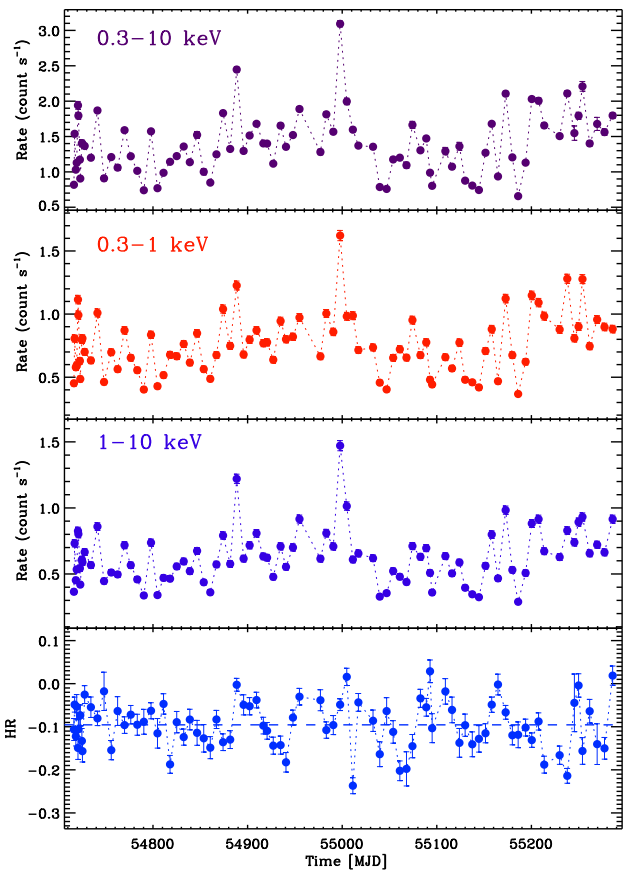}
\caption{Top three panels: total, soft, and hard
XRT light curves of \pks.
Bottom panel: Hardness ratio (HR=(h-s)/(h+s)) light curve; the dashed line represents the 
average value of the hardness ratio. } 
\label{figure:fig1}
\end{center}
\end{figure}
\subsection{X-ray variability}
Figure~\ref{figure:fig1} shows the total (0.3--10 keV), soft (0.3--1 keV), and hard (1--10 keV)
XRT light curves of \pks\ as well as the hardness ratio $HR=(h-s)/(h+s)$ light curve. 
Soft and hard time series appear to vary significantly and roughly in concert. 
According to a  $\chi^2$ test, the count rate variation are highly significant with
$\chi^2/dof=6412/86$ and $5657/86$ for the soft and hard band respectively. 
Based on the same test, the variability
of the $HR$ light curve appears to be statistically significant with $\chi^2/dof=554/86$,
suggesting the presence of spectral variability.
An analysis of the fractional variability $F_{\rm
var}=\sqrt{\sigma^2-\Delta^2}/\langle r\rangle$ (where $\sigma^2$ is the variance, 
$\Delta^2$ the mean square value of the uncertainty associated with each individual count 
rate, and $\langle r\rangle$ the unweighted mean count rate )
confirms the presence of strong variability
at a similar level in the soft and hard X-ray bands with
$F_{\rm var,soft}=(30.6\pm0.4)$\%, $F_{\rm var,hard}=(32.3\pm0.4)$\%. 
\begin{figure}
\begin{center}
\includegraphics[bb=45 32 350 300,clip=,angle=0,width=8.5cm]{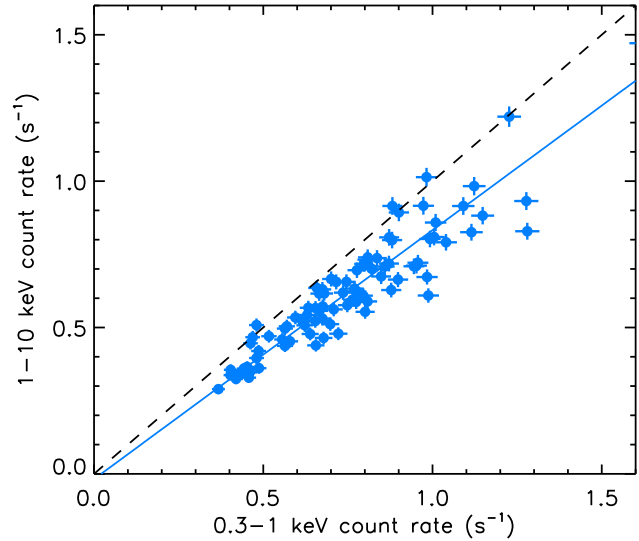}
\caption{Hard vs. soft X-ray count rate plot of \pks\ obtained in the \swift\ XRT campaign.  The 
continuous line represents the best-fit linear model, whereas the 
black dashed line refers to the perfect one-to-one correlation.} 
\label{figure:fig2}
\end{center}
\end{figure}
\begin{figure}
\begin{center}
\includegraphics[bb=80 1 470 645,clip=,angle=0,width=8.5cm]{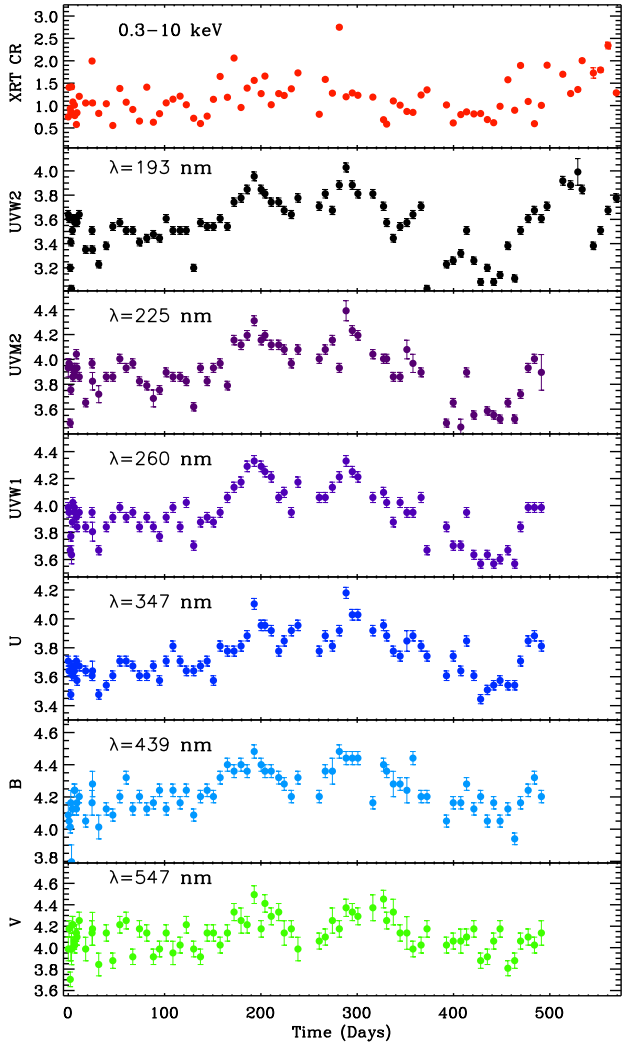}
\caption{\swift\ XRT and UVOT UVW2 light curves of \pks\ from 7 September 2008 to 30 March 2010;
for the other UVOT filters the light curves end on 12 January 2010.
The optical/UV flux densities are corrected for Galactic absorption and expressed in
units of mJy.} 
\label{figure:fig3}
\end{center}
\end{figure}
\begin{figure}
\begin{center}
\includegraphics[bb=20 60 660 490,clip=,angle=0,width=8.5cm]{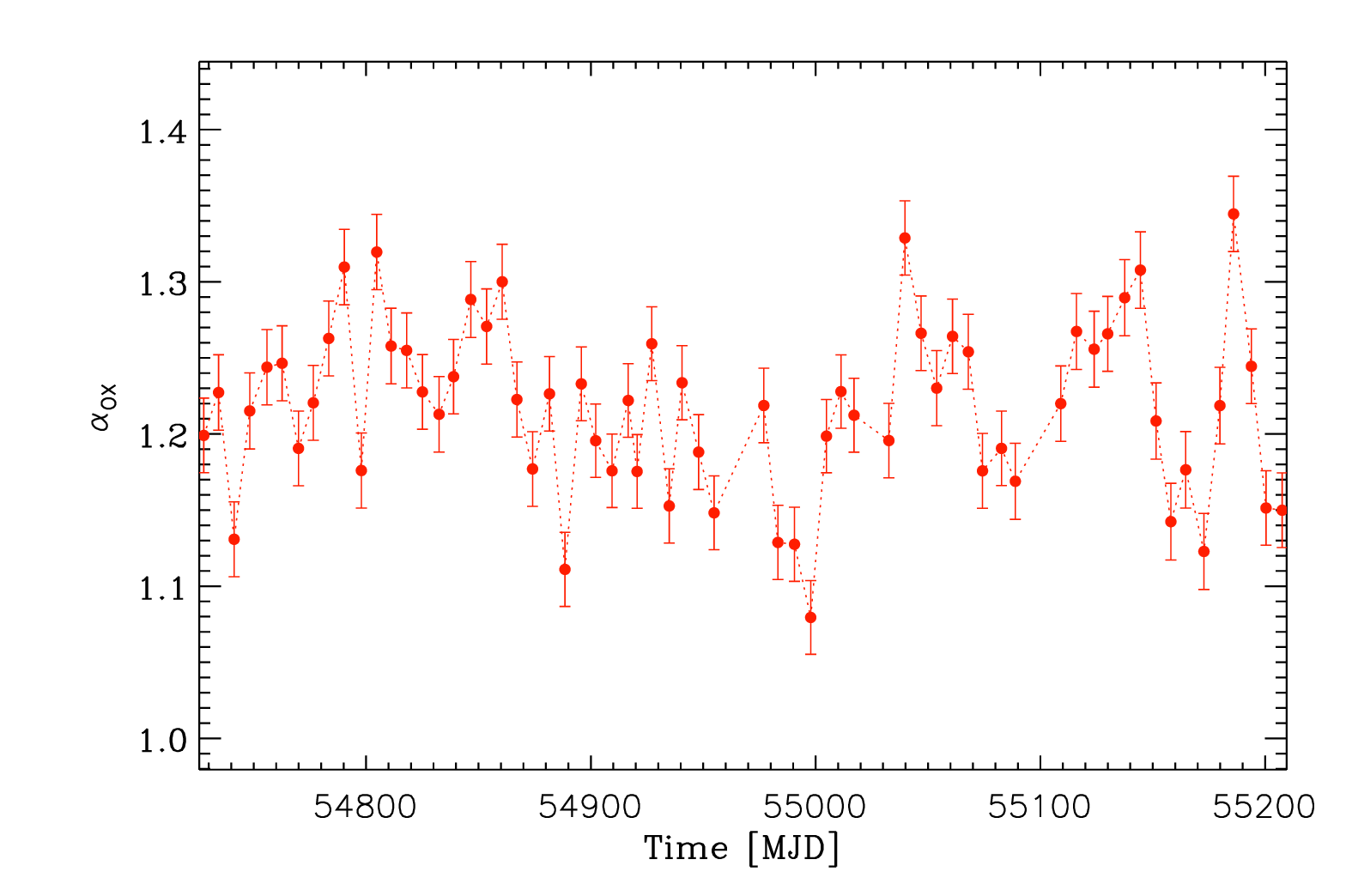}
\caption{Light curve of the broadband spectral index $\alpha_{OX}$.} 
\label{figure:fig4}
\end{center}
\end{figure}

Plotting the X-ray soft count rate versus the hard count rate is a simple
and model-independent way to test for spectral variability \citep[e.g.,][]{chura01}.
A visual inspection of Fig.~\ref{figure:fig2}, which shows the plot of
the soft (0.3--1 keV) and hard (1--10 keV) energy band count rates, suggests the presence of a strong positive correlation.
The dashed line represents the perfect one-to-one correlation ($y=x$), whereas the
continuous line, $y=(-0.02\pm0.01) + (0.85\pm0.02) x$,
 indicates the best linear fit obtained using the routine 
{\tt fitexy} \citep{press97}, which accounts for the errors not only on the 
y-axis but along the x-axis as well, and will be adopted in the rest
of the paper for any linear correlation analysis. 
The best-fit intercept value consistent with zero rules out the 
presence of a non variable X-ray component. The best-fit slope less than 
unity is consistent with a ``softer when brighter" spectral variability trend. 
The non-negligible scatter (especially at higher count 
rates) indicates that, on top of the flux-related spectral variability pattern, 
there also exist additional flux variations of moderate amplitude.

\subsection{UVOT variability}
The light curves obtained from the six UVOT filters (and, for completeness, from 
the XRT) are shown in Fig.~\ref{figure:fig3}. The plotted optical and UV time 
series are the flux densities in units of mJy and have been corrected for Galactic 
absorption.  
All UVOT light curves appear to vary 
substantially (with larger amplitudes at higher energies) on timescales of 
months with two pronounced peaks roughly occurring $\sim$200 and $\sim$300 
days after the beginning of the \swift\ campaign.
According to a  $\chi^2$ test, all UVOT light curves show significant variability
with $\chi^2/dof$ ranging from 3671/76 for the UVW2 filter to 288/76 for the V
filter.
The presence of significant variability at all wavelengths is confirmed by a 
fractional variability analysis that yields: 
$F_{\rm var,V}=(3.1\pm0.3)$\%, $F_{\rm var,B}=(2.9\pm0.1)$\% 
$F_{\rm var,U}=(4.0\pm0.1)$\%, $F_{\rm var,W1}=(4.7\pm0.1)$\% 
$F_{\rm var,M2}=(5.8\pm0.1)$\%, $F_{\rm var,W2}=(6.3\pm0.1)$\%.

Model-independent information on the broadband spectral variability can be
inferred by studying the temporal evolution of the broadband spectral index 
$\alpha_{\rm OX}=
\log(l_{\rm 2500\AA}/l_{\rm 2keV})/\log(\nu_{\rm 2500\AA}/\nu_{\rm 2keV})$,
where $l_{\rm 2keV}$ and $l_{\rm 2500\AA}$ are the monochromatic luminosities
corrected for Galactic absorption
\citep{tanan79}. We derived $\alpha_{\rm OX}$ from the simultaneous X-ray and 
UVW1 filter fluxes, and plotted the light curve in Figure~\ref{figure:fig4},
which confirms the presence of long-term variability of the SED: $F_{\rm var,\alpha_{\rm OX}}=(4.1\pm0.3)$\%.

In summary, the long-term \swift\ campaign of \pks\ confirms the presence of significant
large-amplitude variability in all bands probed by the UVOT and XRT, with the
X-ray band being by far the most variable ($F_{\rm var,X-ray}\sim30$\%) followed by
the UV band ($F_{\rm var,W2}\sim 6$\%) and finally the optical band
($F_{\rm var,V}\sim3$\%).

\section{Inter-band Correlation analysis}
\begin{figure*}
\begin{center}
\includegraphics[bb=40 275 550 570,clip=,angle=0,width=17.cm]{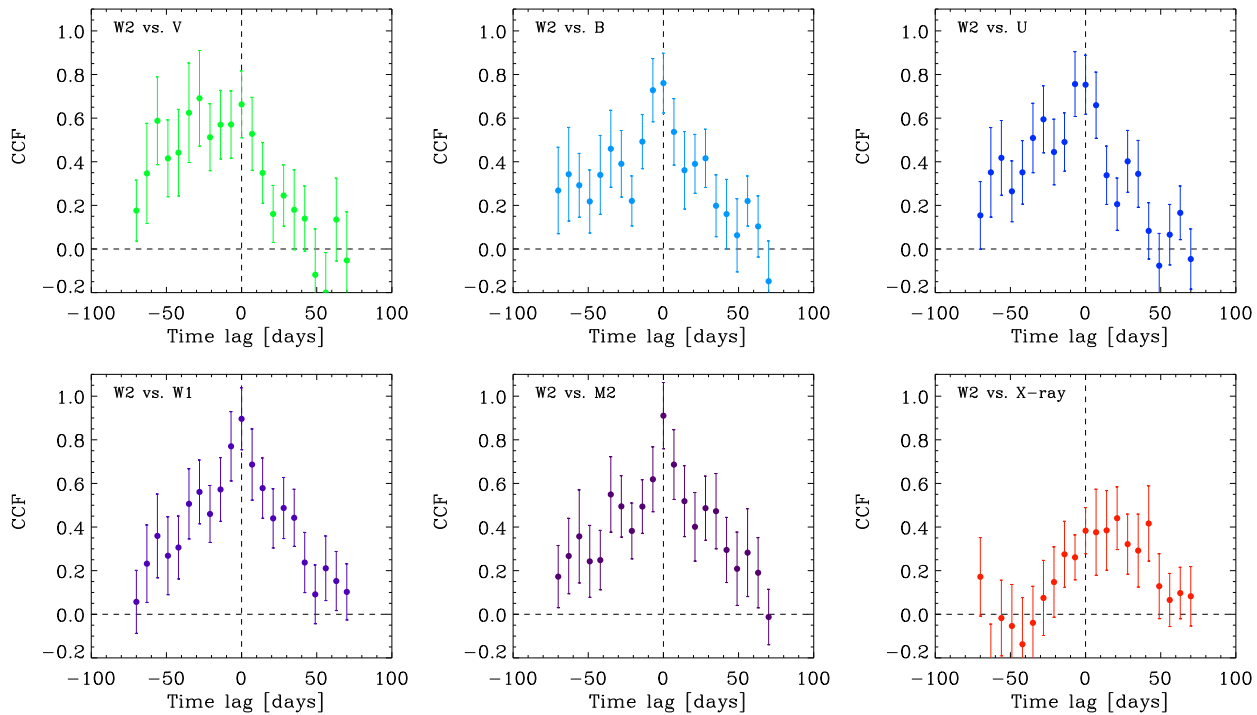}
\caption{Plots of cross-correlation between the UVW2 flux and the other UVOT
filters as well
as the count rate in the 0.3--10 keV energy band.} 
\label{figure:fig5}
\end{center}
\end{figure*}
Important physical insights into the nature of the central engine of \pks\ may be 
obtained from the time delay between the changes occurring in the different energy 
bands probed by the XRT and UVOT.
To this end, we calculated the cross correlation function (CCF) between the UVW2 flux
(which is the most variable UVOT band and likely tracks the behaviour of the inner disc) 
and the other UVOT filters as well as the count rate in the 0.3--10 keV energy band.

We used the discrete correlation function (DCF) method of \citet{edel88} to compute the correlation function at lags $k=0,\pm l\Delta t$, where $l=1,\dots,10,~ \Delta t=7$ days.
To estimate the DCF at each lag, $k$, we used data points with a time separation of $k\pm 1.5$ days. Positive lags mean that the reference light curve (i.e., UVW2) leads, negative lags indicate that the reference light curve follows. The resulting CCFs are shown in Figure~\ref{figure:fig5}. The CCF amplitude rises from $\sim 0.4$ in the case of the UV/X-ray CCF, to $\sim 0.5-0.7$ and $\sim 0.8-0.9$ in the case of the UV/optical band and UV/UV CCFs, respectively. Interestingly, the CCFs are skewed towards negative lags when UVW2 is correlated with the optical bands. The CCF is roughly symmetric when UVW2 is correlated with the other UV bands, and becomes strongly skewed towards positive lags when correlated with the X-ray light curve. This is an indication that the optical band variations {\it lead} those in the UV band, which in turn lead the variations in the X-ray band. Furthermore,
the correlation is weaker between the UV and the X-ray/optical band light curves, which is not surprising, given the fact that the optical band light curves are much ``smoother" than the UV band light curves, while the opposite is true for the X-ray band light curve (see Fig.~\ref{figure:fig3}). 

In Table 3, in addition to the maximum values of the CCF,  we report the time delays (``Lags") between the various bands along with their respective 68\% errors that were  estimated using the Monte Carlo simulation approach proposed by \citet{peter98}. 
For each light curve pair, that we cross-correlated, we produced 5000 simulated light curves following their ``random subset selection" prescription. We computed the DCF of each ``synthetic" light curve pair, in the same way as we did with the observed light curves. We recorded the time lag, $\tau_{\rm peak}$, with the maximum CCF value, CCF$_{\rm max}$, and then we used all the neighboring time lag values with CCF values larger or equal than 0.75 CCF$_{\rm max}$ to estimate their mean, which we accept as the ``centroid time lag", $\tau_{\rm cent}$ for each simulated light curve pair. We used the 5000 $\tau_{\rm cent}$ values to  build up its distribution function, and we estimated its median (which is what we list as ``Lag" in Table 3) and the 68\% confidence limits.  Although there is not a statistically significant detection of a delay between any of the light curve pairs considered (all the ``lags" listed in Table 3 are consistent with zero within their 90\% confidence limits), the lags appear to be in agreement with the visual inspection of Fig.~\ref{figure:fig5}  that we discussed above. The delays are negative and decrease with decreasing energy ``separation" between the the UVW2 and the other optical and UV band light curves, while the delay becomes positive in the case of the UVW2 and the X-ray band light curve.  This putative trend appears to be at odds with the X-ray reprocessing scenario (where the X-ray flux changes lead the UV and optical variations) and is instead consistent with a scenario where variations are produced in the outer part of the accretion flow and propagate inwards, affecting first the optical, then the UV and lastly the X-ray light curves.

\begin{table} 
\footnotesize
\caption{UVOT Correlation analysis results}
\begin{center}
\begin{tabular}{ccc} 
\hline        
\hline
\noalign{\smallskip}
Energy bands    & Lag (d) & ccf$_{\rm max}$\\
(1) & (2) & (3)\\
\noalign{\smallskip}
\hline 
\noalign{\smallskip}
UVW2 X-ray  &   $16.8^{+14.7}_{-16.8}$ & $0.58^{+0.15}_{-0.14}$\\
\noalign{\smallskip}
\hline
\noalign{\smallskip}
UVW2 UVM2  &    $0.0^{+9.3}_{-15.8}$ & $0.86^{+0.14}_{-0.12}$\\
\noalign{\smallskip}
\hline
\noalign{\smallskip}
UVW2 UVW1  &   $-3.5^{+7}_{-10.5}$ & $0.84^{+0.13}_{-0.10}$ \\ 
\noalign{\smallskip}
\hline
\noalign{\smallskip}
UVW2 U  &   $-8.4^{+8.4}_{-12.6}$ & $0.77^{+0.12}_{-0.10}$ \\
\noalign{\smallskip}
\hline
\noalign{\smallskip}
UVW2 B  &  $-4.7^{+6.6}_{-16.3}$ & $0.76^{+0.13}_{-0.11}$\\
\noalign{\smallskip}
\hline
\noalign{\smallskip}
UVW2 V  &   $-23.8^{+14.5}_{-11.2}$ & $0.80^{+0.19}_{-0.12}$ \\
\noalign{\smallskip}
\hline        
\hline
\end{tabular}
\end{center}
{\bf Columns Table 3}: 1= Correlated energy bands. 
2= Lags measured in days with the 68\% errors. 3= Maximum of CCF with the 68\% errors.
\label{tab3}
\footnotesize
\end{table}  

\section{Broadband Spectral Analysis}
In this section, we investigate the evolution of the 
broadband SED of \pks\ over 1.5 years.
For this purpose, we follow the procedure adopted in our previous work on
the short-term evolution of the SED of \pks\ and utilize
a spectral model that comprises a disc and two Comptonisation components, 
{\tt discPN+WABS(BMC+BMC)}, which yielded adequate fits of the SEDs 
obtained from contemporaneous observations of \swift\ and \xmm\
in September 2007 \citep{glioz10}. 
The absorption model {\tt WABS}, with $N_{\rm H}$ fixed at the Galactic value, is 
only applied to the Comptonisation models because the optical and UV data,
which are fitted with  {\tt discPN}, have already been corrected for Galactic absorption. 
\begin{figure}
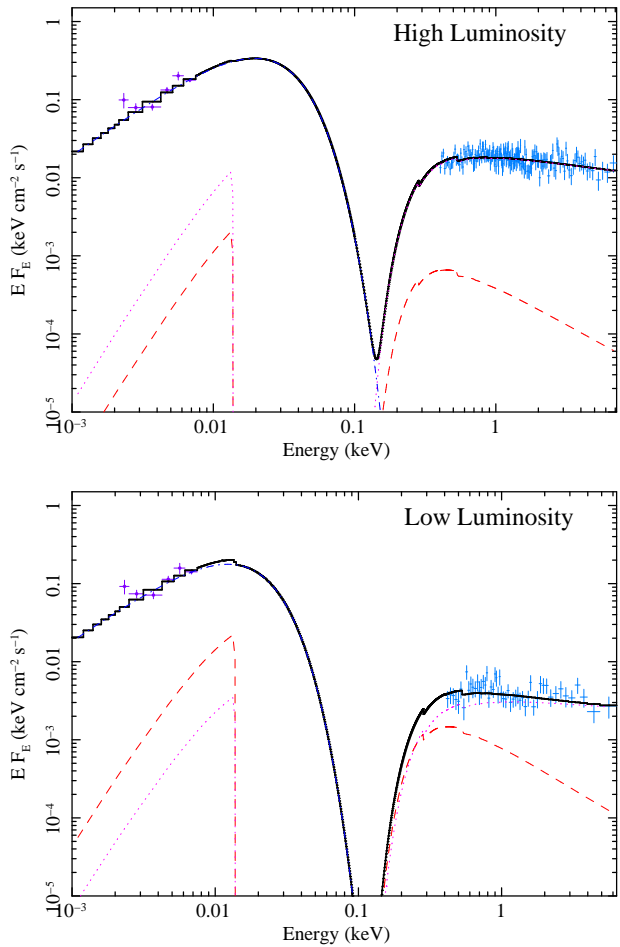

\includegraphics[bb=60 10 595 720,clip=,angle=-90,width=8.5cm]{fig6a.eps}

\includegraphics[bb=60 10 595 720,clip=,angle=-90,width=8.5cm]{fig6b.eps}

\caption{Deconvolved XRT spectra in the 0.3--10 keV energy range,
combined with UVOT data. The overall SEDs are fitted with a discpn model and two BMC models to characterize the X-ray spectrum. The dotted and dashed lines describe the BMC2 and BMC1 models, respectively; the
diskpn line is indistinguishable from the continuous line that represents the resultant model.
}
\label{figure:fig6}
\end{figure}

The {\tt discPN} model \citep{gierl99} is a generalization of the disc
black body which includes corrections for the temperature distribution near the black 
hole. This model has three parameters: the maximum disk temperature $kT_d$, 
the inner radius $R_{\rm in}$, and the 
normalization $N_d$, which depends on the black hole mass, the distance, the inclination 
angle $i$, and the colour factor $\beta$. 
The {\tt BMC} model \citep{tita97} is  a simple and robust Comptonisation model 
that can describe both thermal and bulk Comptonisation processes; it is
characterized by four parameters: the temperature of the
thermal seed photons $kT$, the energy spectral index $\alpha$ 
(where $f_\nu\propto \nu^{-\alpha}$), a parameter 
$\log(A)$ related to the Comptonisation fraction $f$ (i.e., the ratio
between the number of Compton scattered photons and the number of seed 
photons) by the relation
$f=A/(1+A)$, and the normalization $N_{\rm BMC}=L_{39}/d_{10}^2$, where 
$L_{39}$ is the luminosity in units of $10^{39}{~\rm erg~s^{-1}}$ and $d$ is
the distance in units of 10 kpc. 

First, we fitted the UVOT data and the hard X-rays ($E>1$ keV) with the discPN model and
one BMC model only (hereafter $BMC2$), keeping $R_{\rm in}$ fixed at $6~R_{\rm G}$ and leaving the
other parameters ($N_d$, $kT_d$, $\alpha_2$, $\log(A_2)$, $kT_2$, and $N_{\rm BMC2}$) 
free to vary. We also kept the temperature of seed photons of the Comptonisation model 
fixed to the maximum temperature of the accretion disc, since we assume that the seed photons
are produced by the disc.
Then, we considered the 0.3--1 keV soft X-ray data and added another BMC component (hereafter $BMC1$),
keeping the $BMC2$ parameters fixed at their best fit values (except for $N_{\rm BMC2}$
which is free to vary). We found that $N_{\rm BMC1}$ was significantly different from zero in the
vast majority of the cases. This result indicates that a ``soft excess" (i.e., a significant
flux in excess to the extrapolation of the BMC2 component to energies $<$ 1 keV) is present
in nearly all the XRT observations. This is a well known result for \pks, as 
this soft excess has been detected in all previous observations with all the major X-ray
satellites in the last two decades \citep[e.g.,][]{lei99,obri01,brink04}.


All SEDs are reasonably well fitted with this model: for all broadband spectra 
the reduced $\chi^2$ values
range between 0.75 and 1.54 with a mean value of 1.10 and a standard deviation
of 0.16 for an average number of degrees of freedom of 95.
Two examples of SEDs fitted with the model described above, that represent 
the highest (MJD=54997.82) and lowest (MJD=55144.51) luminosity  occurrences
of \pks\ during the \swift\ campaign, are shown in Figure \ref{figure:fig6}.
Both SEDs are well fitted by this model ($\chi^2_{\rm red}$=1.18 and 1.28
for high-$L$ and low-$L$, respectively), they are 
clearly dominated by the disc emission and 
have similar shapes: the spectral parameters are consistent within their 
uncertainties, with $kT\sim 7-8$ eV, $\Gamma_1\sim 3-5$, $\log(A_1)\sim 2$,  
$\Gamma_2\sim 1.8-2.2$, $\log(A_2)\sim 0.5-1$. 
Similar values are found for all SEDs. More specifically, the temperature 
of the disc appears to be fairly stable, whereas
the disc normalization changes considerably (see below for more details). 
For BMC2 (the Comptonisation model
parametrizing the hard part of the X-ray spectrum) $\Gamma_2$ ranges
between 1.6 and 2.4, and $\log(A_2)$ between 0.5 and 2. For BMC1, all the
parameters are poorly constrained with  $\Gamma_1$ consistently steep ($>3$)
and $\log(A_1)$ that shows systematically high values and hence was fixed at 2.


\begin{figure}
\begin{center}
\includegraphics[bb=40 3 540 570,clip=,angle=0,width=8.5cm]{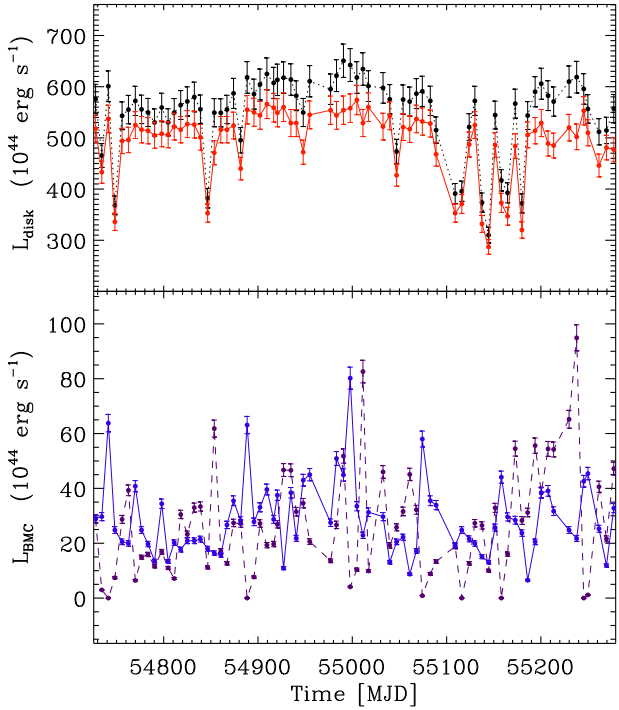}
\caption{Long-term evolution of the spectral components used to fit the simultaneous SEDs
produced by the \swift\ XRT and UVOT. In the top panel, the dotted black line follows the
trend of the total luminosity and the continuous (red) line indicates $L_{\rm disc}$. In the
bottom panel, the continuous line represents $L_{\rm BMC1}$ and the dashed one indicates
$L_{\rm BMC2}$.} 
\label{figure:fig7}
\end{center}
\end{figure}
Figure \ref{figure:fig7} shows the evolution of the luminosity values (in units of $10^{44}{~\rm erg~s^{-1}}$) associated with the different spectral components. 
The top panel
represents the temporal evolution of the disc luminosity (continuous line),
which has a mean value of $4.9\times 10^{46}{~\rm erg~s^{-1}}$,
and of the total luminosity (dotted line, 
$\langle L_{\rm tot} \rangle=5.5\times 10^{46}{~\rm erg~s^{-1}}$) that is 
obtained by integrating the best fit model between 0.001 and 10 keV.
The close similarity between the two patterns and the consistency between
the respective values of fractional variability 
($F_{\rm var,L_{\rm tot}}= F_{\rm var,L_{\rm disc}}\simeq 12\%$)
confirms that total luminosity changes are dominated by $L_{\rm disc}$.
The bottom panel shows the temporal evolution of the luminosities
associated with the two BMC components (obtained by their integration over the 
0.1--10 keV energy range; $\langle L_{\rm BMC1} \rangle=2.6\times 10^{45}{~\rm erg~s^{-1}}$,
$\langle L_{\rm BMC2} \rangle=2.9\times 10^{45}{~\rm erg~s^{-1}}$). Both light curves are characterized by continuous large-amplitude changes, which do not appear to correlate.
$L_{\rm disc}$ and $L_{\rm BMC2}$ are well determined, but the decomposition between
$BMC1$ and $BMC2$ may not be very accurate, since the latter extends to energy below 1 keV. 
As a consequence,  it is the 0.3--1 keV flux and not the $BMC1$ flux that better  
reflects the soft excess contribution.

\begin{figure}
\begin{center}
\includegraphics[bb=30 10 540 570,clip=,angle=0,width=8.5cm]{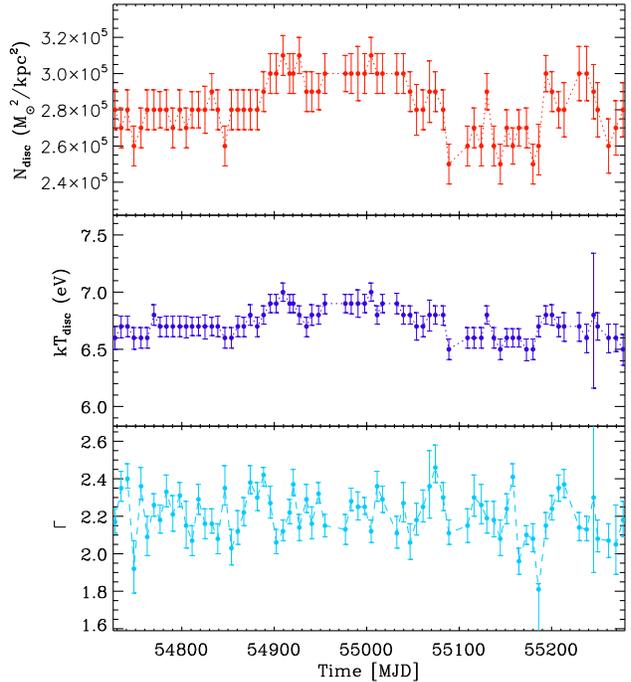}
\caption{Long-term evolution of the spectral parameters $N_d$ (top panel),
$kT_d$ (middle panel), and the hard photon index (bottom panel).} 
\label{figure:fig8}
\end{center}
\end{figure}

Due to the short exposures (and hence low S/N in the spectra) inherent in the 
monitoring approach,
the best fit parameter values of the disc and Comptonization components 
are not well constrained. As a result, despite the fact that the disc, BMC1 and BMC2 
luminosities are clearly variable, we cannot determine which model parameter 
is the main driver of these variations, as all model parameters are consistent 
with the hypothesis of being constant (due to the large errors associated with them). 
For this reason, we repeated the model fitting procedure of the broadband SEDs twice. 
In the first case, we kept $kT_d$ fixed to the mean of the best-fit values from the 
previous fit (which is equal to 6.8 eV), and let all the model parameters free 
to vary following the fitting procedure described above. In the second case, we 
kept $N_d$ fixed 
to the mean of the best-fit values from the previous fit (which is equal to 
$2.82\times 10^5~{\rm M_\odot^2/kpc^2}$).
The results of this spectral fitting procedure
are shown in Fig. \ref{figure:fig8}, where, for comparison, we also plot the hard 
X-ray photon index in the bottom panel. According to a $\chi^2$ test,  $kT_d$,
$N_d$, and $\Gamma_2$ appear to be variable, however, a fractional variability
analysis ($F_{\rm var}=0.7\pm0.4\%, 3.6\pm0.6\%, 3.6\pm0.7 \%$, respectively)  
favors $N_d$ as the main driver of the disc luminosity.

\begin{figure*}
\begin{center}
\includegraphics[bb=40 340 590 500,clip=,angle=0,width=17cm]{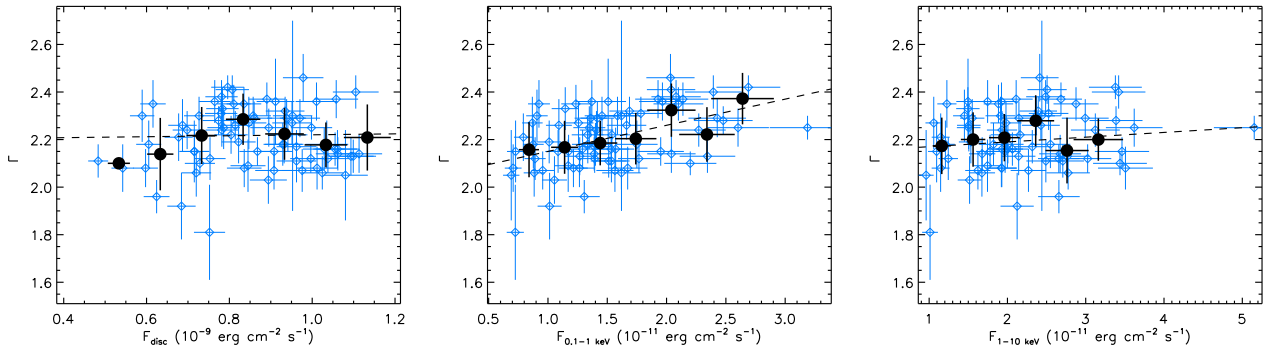}
\caption{2--10 keV photon index $\Gamma$
plotted versus the disc flux (left panel), the soft-excess (middle panel)
flux, and the hard X-ray flux (right panel). The filled circles represent the
binned data and the dashed line represents the best-fit linear model.}
\label{figure:fig9}
\end{center}
\end{figure*}

To shed some light on the link between disc and corona and specifically on the 
origin of
the seed photons that are subsequently Comptonised by the corona, we have studied
the correlation between the hard photon index (which parameterizes
the coronal activity) and three different fluxes: the disc flux, the soft-excess
flux, and the hard X-ray flux. The results of this analysis, performed keeping $kT_d$
fixed to the mean value obtained from all the best fits, are shown in Figure \ref{figure:fig9}.
A visual inspection of this figure reveals that only the flux associated with the
soft-excess shows a clear positive trend. This finding is confirmed by the
formal fitting of the data, $\Gamma=(2.02\pm0.04) + (0.12\pm0.02)F_{\rm 0.3-1~keV}$,
as well as by the Spearman's and Kendalls's rank correlation 
coefficients ($\rho=0.42$ and $\tau=0.30$, respectively) and their 
associated chance probabilities ($P<2\times10^{-4}$). On the other  hand,
no statistically significant correlation is found for $\Gamma$ vs. $F_{\rm disc}$
($\Gamma=(2.20\pm0.06) + (0.02\pm0.07)F_{\rm disc}$, $\rho=-0.003$ and $\tau=-0.01$),
or for $\Gamma$ vs. $F_{\rm Hard~X-rays}$
($\Gamma=(2.16\pm0.04) + (0.02\pm0.01)F_{\rm 1-10~keV}$, $\rho=0.11$ and $\tau=0.08$).

In conclusion, the overall SEDs of \pks\  are
adequately fitted by a disc model parameterizing the optical/UV emission and two 
Comptonisation components that respectively describe the soft excess and coronal emission. 
The SED is dominated by a variable disc component
($3\times10^{46}{~\rm erg~s^{-1}}<L_{\rm disc}<6\times10^{46}{~\rm erg~s^{-1}}$),
with substantial contribution from the two highly variable Comptonisation components
($1\times10^{44}{~\rm erg~s^{-1}}<L_{\rm BMC1}<9\times10^{45}{~\rm erg~s^{-1}}$
and 
$7\times10^{44}{~\rm erg~s^{-1}}<L_{\rm BMC2}<8\times10^{45}{~\rm erg~s^{-1}}$).
The coronal changes parameterized by the hard X-ray photon index appear to be
correlated
with the soft excess flux suggesting that the latter component is the source of
the seed photons.

\section{Discussion and Conclusion} 
To put in context the main findings of our 1.5 year \swift\ monitoring campaign of \pks,
we will summarize the most relevant observational results and discuss their implications.

\begin{itemize}
\item {\it X-ray flux and spectral variability:}
Throughout the 1.5 year \swift\ XRT monitoring campaign, \pks\ showed continuous 
large-amplitude flux variations with similar variability levels in the soft and 
hard X-ray band: $F_{\rm var,0.3-1~keV}\simeq F_{\rm var,1-10~keV}\sim30$\%.
Analogous values of $F_{\rm var}$ were measured in the
1-year \rxte\ monitoring between March 2005 and March 2006, suggesting that
the variability level of \pks\ is considerably stable over a time interval
of several years and an energy band ranging from 0.3 to 15 keV.

Model-independent analyses based on long-term XRT data revealed that soft 
(0.3--1 keV) and hard (1--10 keV) X-ray count rates appear to broadly vary in 
concert, as indicated by the similar variability trends in Fig. \ref{figure:fig1}
and by the tight correlation in the soft-hard count rate plot (see Fig. \ref{figure:fig2}).
The apparent variability of the $HR$ light curve (shown in the bottom panel of 
Fig. \ref{figure:fig1} and confirmed by a $\chi^2$ test) 
and the substantial scatter in the soft-hard count rate plot indicate the presence of
significant spectral variability in the 0.3--10 keV band, which can be explained by the
presence of a second component in the X-ray band (i.e., the soft excess) whose variability
properties are not identical to those of the hard X-ray component.

The results from the broad-band fits to the optical/UV/X-ray spectra indicate significant
spectral variability within the 1--10 keV, as indicated by the $\Gamma$
variations in Fig.\ref{figure:fig8}. Furthermore,
the plot of $\Gamma$ vs. soft X-ray flux (see the middle panel of Fig. \ref{figure:fig9}), reveals the existence of a positive correlation and suggests
that the soft excess is the source of seed photons for the corona. In this scenario,
the soft excess  may physically represent a ``hot skin" on top of the disc,
which is adequately fitted by a low-temperature Comptonisation component, in
agreement with previous 
spectral analysis based on higher quality X-ray spectra  from \xmm\ 
\citep{brink04,papa10a}. Correlations between soft photons and hard photon index 
represent one of the strongest evidence in favor of Comptonisation in radio-quiet
AGN, as demonstrated by recent work on different AGN
\citep[see, e.g.,][]{nan00,done12,petru13}

On the other hand, there is no clear evidence for a strong positive correlation
between $\Gamma$ and hard X-ray flux (see Fig. \ref{figure:fig9} right panel),
which is typically observed in Seyfert galaxies when $\Gamma$ is plotted versus 
the hard X-ray flux \citep[e.g.,][]{papa02,sobo09}. However, it must be kept in mind
that \pks\  accretes at a higher level (based on the \swift\ data \pks\ appear to 
accrete at super-Eddington rate with $L_{\rm bol}/L_{\rm Edd}\sim 1-2$,
 where the range encompasses the variability of the integrated $L_{\rm bol}$ and the
uncertainty on the black hole mass)
compared to typical Seyfert galaxies. Interestingly, this spectral behaviour is 
consistent with our
findings based on the long-term \rxte\ campaign of \pks\ in the 2--15 keV band
which we interpreted as an indication that \pks\ is a large-scale analog of
GBHs in highly-accreting intermediate spectral state.

\item {\it Optical and UV variability:}
Although previous studies revealed the UV dominance of the SED  of \pks\ 
\citep[e.g.,][]{obri01}, the brevity of the observations hampered a
meaningful study of the UV variability. The long-term monitoring with the
\swift\ UVOT made it possible for the first time to investigate 
simultaneously the variability in the optical and UV bands, which are crucial 
to probe the activity of the accretion disc. Our 1.5 year \swift\ campaign
revealed that \pks\ is significantly variable in all the energy bands covered 
by the UVOT with moderately large amplitude variations occurring on timescales
of months with variability levels increasing from the reddest 
($F_{\rm var,V}\sim3$\%) to the bluest filter ($F_{\rm var,W2}\sim 6$\%).
The UVOT variability is substantially
lower than the X-ray variability, in agreement with the findings of \citet{grupe10},
who studied the variability of the largest sample of AGN with simultaneous
coverage of the UV and X-ray energy ranges and found that UV emission only varies 
marginally. 

The disc component is reasonably well fitted by a discPN model and the $L_{\rm disc}$ 
variability appears to be dominated by variations of the disc normalization, which 
depends on physical properties of the system that are constant (such as the distance,
the black hole mass, and the inclination angle) and on the colour factor, whose
temporal evolution is poorly understood. The limited number of data points in the
optical-UV bands do not allow to use more sophisticated disc models that may provide
more direct clues on the dominant variable physical parameters of the accretion flow.
Nevertheless, following \citet{dext11}, we can speculate that the disc variations 
observed in \pks\ can be explained by local $kT$ fluctuations in a strongly 
inhomogeneous disc.

A model-independent analysis based on the light curve of $\alpha_{\rm OX}$ 
confirms the presence of persistent broadband spectral variability with
$\alpha_{\rm OX}$ ranging between 1.10 and 1.35. The existence
of significant intrinsic variability, $F_{\rm var,\alpha_{\rm OX}}=(4.1\pm0.3)$\%.
is in agreement with recent findings from \citet{van13} that were derived from a
 systematic study of the $\alpha_{\rm OX}$ variability of the bright, soft X-ray 
selected sample of AGN observed with \swift\ \citep{grupe10}.
Combining the variability range of  $\alpha_{\rm OX}$
with the fact that throughout the monitoring campaign the hard X-ray photon 
index varies in the 1.8--2.4 range, makes it possible to locate \pks\
in the region of the $\Gamma-\alpha_{\rm OX}$ diagram consistent with GBHs
in the intermediate state (see Figure 6 of \citealt{sobo09b}).

\item {\it Correlation analysis:}
A combined study of the \swift\ UVOT and XRT light curves of \pks\ indicates 
that the UVW2 data are highly correlated with all the other UVOT bands and 
weakly correlated with the XRT.
A CCF analysis confirms these results and suggests the presence of an evolution 
trend in the time lags, although not at a statistically significant level.
More specifically, the changes in the optical bands appear to lead those
in the UV bands, which in turn lead the X-ray variations. 

These results do not support the ``reprocessing model", where the UV variable component
is thought to be caused by X-ray irradiation from the corona. Indeed, several
past studies have favored this model for different AGN
(e.g., Mrk~79, \citealt{bree09}; NGC~3783, \citealt{arev09}; NGC~4051, 
\citealt{bree10}), including a recent work based on the
long-term \swift\ monitoring of the low mass AGN, NGC~4395 \citep{came12}.
However, in the reprocessing scenario X-ray changes are predicted to lead 
the UV, which in turn should
lead the optical band variation, which is at odds with the observational
results of \pks. 

On the other hand,
this trend appears to be consistent with a scenario where perturbations 
rapidly propagate from the outer part of the accretion flow (responsible for the
optical emission) to the inner hotter region (that emits UV and is closely connected  
to the X-ray emitting corona). A possible problem with this interpretation is that the
time lags inferred from the CCF analysis (see Table 1) appear to be
significantly shorter than the predicted propagation timescales
in a standard accretion disc \citep[e.g.,][]{frank02}.
However, the apparent discrepancy between measured time lags and predicted 
propagation timescales can be reconciled considering that \pks\ is likely 
accreting at or above the Eddington rate and therefore its accretion flow  
cannot be parametrized by a standard accretion disc but rather by a slim
disc whose typical timescales are considerably shortened \citep[e.g.,][]{kawa03}.

Another possible problem for the propagation model is the amplitude of the variations:
the fractional variability measured in the UV bands is larger than the one measured  in 
the optical bands both in percentage and using physical units (i.e., multiplying 
$F_{\rm var}$ by the average luminosity in each specific energy band:
$F_{\rm var,L_V}=(3.2\pm0.3)\times 10^{43}{~\rm erg~s^{-1}}$, 
$F_{\rm var,L_B}=(3.6\pm0.1)\times 10^{43}{~\rm erg~s^{-1}}$,
$F_{\rm var,L_U}=(6.4\pm0.2)\times 10^{43}{~\rm erg~s^{-1}}$, 
$F_{\rm var,L_{W1}}=(8.1\pm0.2)\times 10^{43}{~\rm erg~s^{-1}}$, 
$F_{\rm var,L_{M2}}=(13.2\pm0.3)\times 10^{43}{~\rm erg~s^{-1}}$, 
$F_{\rm var,L_{W2}}=(15.3\pm0.2)\times 10^{43}{~\rm erg~s^{-1}}$).
However, the emission from the inner disc regions can be additionally affected by
localized variations which do not influence the emission of the
optical band produced at larger radii. If these putative variations are of larger
amplitude (which is plausible, since most of the energy is released in the inner part of
the disc), then the different levels of variability of \pks\ measured by the UVOT can
be reconciled  with the inward propagation scenario. Furthermore, these extra variations
may also explain the fact that the correlation between the UV and the optical bands
is not as strong as the correlation among UV bands. This scenario can also explain
the observed delay of X-ray light curves with respect to the UV.

\end{itemize}

In conclusion, our study confirms that \pks\ is a highly accreting system,
whose optical, UV, and X-ray radiation is dominated by accretion-related 
emission. On long timescales, \pks\ is highly variable at all 
wavelengths probed by the \swift\
UVOT and XRT with variability levels decreasing from the X-rays 
($F_{\rm var,X-ray}\sim30$\%), to the UV ($F_{\rm var,W2}\sim 6$\%), to 
the optical band ($F_{\rm var,V}\sim 3$\%). The large-amplitude variations
measured by the bluest UV filter (UVW2) are highly correlated with all 
the other UVOT bands and weakly (but significantly) correlated with the
X-ray variations, implying a physical link between disc and corona. The CCF
results provide suggestive evidence for perturbations that
rapidly propagate from the outer to the inner parts of the accretion flow
and to the corona, favoring the inward propagation scenario over the 
reprocessing one. However, the latter cannot be completely ruled out, since it
may occur and dominate on timescales shorter than those probed in our long-term
campaign with weekly cadence.
Finally, 
the positive correlation between the soft X-ray flux 
and the hard photon index suggests that in PKS 0558­-504 the seed photons are
provided to the corona by the soft excess component, which may represent the
``hot skin" of the accretion disk.

This work demonstrates that the simultaneous investigation of 
UV, optical, and X-ray variability in AGN is one of the most
effective tools to shed light on their central engine. Further progress
can be obtained by targeting highly variable, clean systems (i.e., without prominent
intrinsic absorption, warm absorbers, and jets), whose $M_{\rm BH}$ and
$\dot m$ are well constrained. \\

\noindent{\bf \Large Acknowledgements}\\ 
MG acknowledges support by the SWIFT Guest Investigator Program
under NASA grant 201564. DG acknowledges support from the NASA Swift program
through contract NAS5-00136.


\end{document}